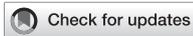

# Structural evolution of international crop trade networks

Yin-Ting Zhang[1] and Wei-Xing Zhou[1,2,3]*

[1]School of Business, East China University of Science and Technology, Shanghai, China, [2]Research Center for Econophysics, East China University of Science and Technology, Shanghai, China, [3]Department of Mathematics, East China University of Science and Technology, Shanghai, China

Food security is a critical issue closely linked to human being. With the increasing demand for food, international trade has become the main access to supplementing domestic food shortages, which not only alleviates local food shocks, but also exposes economies to global food crises. In this paper, we construct four temporal international crop trade networks (iCTNs) based on trade values of maize, rice, soybean and wheat, and describe the structural evolution of different iCTNs from 1993 to 2018. We find that the size of all the four iCTNs expanded from 1993 to 2018 with more participants and larger trade values. Our results show that the iCTNs not only become tighter according to the increasing in network density and clustering coefficient, but also get more similar. We also find that the iCTNs are not always disassortative, unlike the world cereal trade networks and other international commodity trade networks. The degree assortative coefficients depend on degree directions and crop types. The analysis about assortativity also indicates that economies with high out-degree tend to connect with economies with low in-degree and low out-degree. Additionally, we compare the structure of the four iCTNs to enhance our understanding of the international food trade system. Although the overall evolutionary patterns of different iCTNs are similar, some crops exhibit idiosyncratic trade patterns. It highlights the need to consider different crop networks' idiosyncratic features while making food policies. Our findings about the dynamics of the iCTNs play an important role in understanding vulnerabilities in the global food system.

# 1 Introduction

Food security is a mainstay in national security and has become one of the global hot spots [1]. Due to the impact of the COVID-19 pandemic, the number of global hungry people continued to rise in 2020, from 2.05 billion to 2.37 billion, and about 30 million in 2030 more than if the pandemic had not happened[1]. Food supplies face unknown potential risks, with factors such as global pandemics, climate extremes, conflicts and so on. Globalization confers pros and cons with regard to food security [2], providing access to international food trade [3]. On the one hand, international trade meets food demand of some economies with food shortages by supplying food produced elsewhere beyond self-consumption and strategic reserves [4]. On the other hand, trade might multiply disruption to food supply chains [5] and exacerbate economies' vulnerability to sudden shock in global food system [6]. Therefore, international food trade has a crucial impact on food security [7, 8].

Before evaluating underlying benefits and risks for food security, it is necessary to explore characteristics of international food trade. Recently, complex networks have become an important method to study trade relationships existing between pairs of economies in the world [9]. Therefore, many studies have contributed insights into the structure and dynamics of the global food trade system based on network science [10–12]. Some literature focused on one major crop feeding a large number of population, such as maize [13] and wheat [14]. These studies described the trade patterns of international crop trade system [13], and explored the factors that impact the food supply [15]. Investigating the international virtual water network (iVWN) is another common approach to understand global food security [16]. By quantifying water embodied in several food commodities, researchers link the properties of the iVTW to the resilience of the global food system to shocks [17, 18]. However, there are many different definition of resilience [4, 19], or indicators measuring network vulnerability [15, 20].

The topological properties of the international food trade networks (iFTNs) are closely related to the assessment of global food security and should be investigated carefully. Previous literature has focused on the complexity of iFTNs [11] and studied the impact of shocks to the iFTN [21, 22]. However, the microstructure of the iFTN is worth discussing and studying [23]. The evolution of the international food trade system and comparison of trade patterns between different crops still remain a spectrum of investigation. Here, we consider four dietary staples (maize, rice, soybean and wheat), which make up more than 75% of the calories consumed by populations and animals [2, 11]. We construct four international crop trade networks (iCTNs) and quantify the evolution of these iCTNs from 1993 to 2018. Although our work does not specifically evaluate shocks or food security, the dynamics of four different iCTNs provide basic understanding of the global food trade system. The evolution of network features shows the change of iCTNs and the necessity of new methods measuring food security.

In this paper, we attempt to explore and compare the main stylized characteristics pertaining to crop trade relationships and their evolution over time. We focus on structural characteristics such as node degrees, node strengths, link weights, density, the clustering coefficient, reciprocity and assortativity. Our study answers two questions: 1) How has the structure of iCTNs changed over time? 2) What are the differences in trade patterns between different crops? The remainder of this paper is organized as follows. Section 2 describes the data sets used in our work and the construction of the iCTNs. Section 3 presents the empirical results about the dynamics of the four iCTNs. We summarize our results in Section 4.

# 2 Data and method

## 2.1 Data description

We obtained the FAOSTAT data sets on international trade flows from the Food and Agriculture Organization (FAO, http://www.fao.org), which contain the annual bilateral export-import data during the period 1993–2018. The Soviet Union collapsed in 1991 and the world pattern changed dramatically, and Yugoslavia and Czechoslovakia also disintegrated one after another in 1992. Therefore, our data began in 1992 [17]. Since the data sets contain some inconsistencies between the declaration of importers and exporters, we first complied the crop trade matrix by using the import data, and then used the export data to fill data gaps. We got four crop trade matrices $W^{crop}(t)$, and denoted them with superscripts $crop \in$ {M, R, S, W} for maize, rice, soybean and wheat. The number of economies changes as the evolution of political boundaries over time. However, this fact does not affect our analysis of iCTNs [24]. We excluded economies from the annual network analysis when their aggregated values of any kind of crop trade was zero. The final data sets for the network analysis covered 246 economies over the period from 1993 to 2018.

## 2.2 Network construction

We constructed the temporal iCTNs with respect to different crops. The annual iCTN in each year is a multi-layer network, where the nodes represent economies connected by multiple directed links (or links). The link weight $w_{ij}^{crop}(t)$ for a crop is the exports from the economy $i$ to the economy $j$ in a network $G^{crop}(t) = (\mathcal{V}^{crop}(t), W^{crop}(t))$, where $\mathcal{V}^{crop}(t)$ is the set of nodes (that is, the set of economies involved in the trade of *crop* in year *t*). The total networks over 26 years include 246 economies. Not all economies engaged in crops trade in each year, so usually $N_\mathcal{V}^{crop}(t) < 246$. We obtained 26 × 4 yearly international crop trade networks and explored the evolution of structural properties for each iCTN in this work.

---

1 The State of Food Security and Nutrition in the World (2020), available at https://www.fao.org.

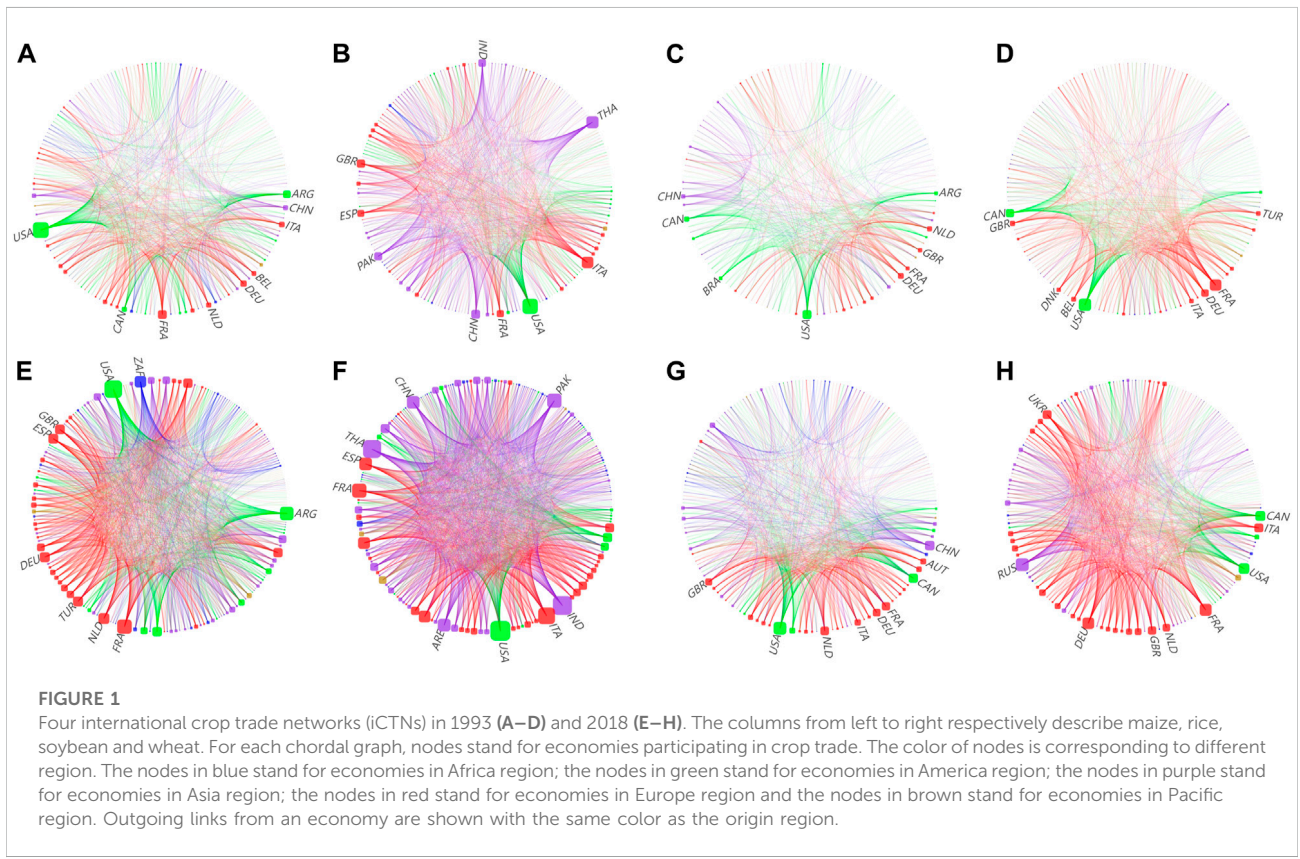

**FIGURE 1**
Four international crop trade networks (iCTNs) in 1993 **(A–D)** and 2018 **(E–H)**. The columns from left to right respectively describe maize, rice, soybean and wheat. For each chordal graph, nodes stand for economies participating in crop trade. The color of nodes is corresponding to different region. The nodes in blue stand for economies in Africa region; the nodes in green stand for economies in America region; the nodes in purple stand for economies in Asia region; the nodes in red stand for economies in Europe region and the nodes in brown stand for economies in Pacific region. Outgoing links from an economy are shown with the same color as the origin region.

Figure 1 shows the four iCTNs in 1993 and 2018. For each economy (node), the symbol size represents its total export value. The thickness of a link represents the trade flow between two economies. It is evident that compared with the iCTNs of 1993, there were more links in the iCTNs of 2018, which indicates that new trade relationships were formed. The nodes became larger and the links became broader, corresponding to larger trade volumes. We note that economies in Asia and Europe are major exporters, especially for maize and wheat. What's more, the United States and Germany are the most important crop exporters that had very large export values in 2018.

## 3 Empirical results

### 3.1 Summary statistics

For each year, we computed network statistics and described the evolution of the four iCTNs. For simplicity,

we omitted the superscript *crop* in the following description. The number of nodes $N_\mathcal{V}$ measures how many economies engaged in trade, and the number of links $N_\mathcal{E}$ measures the trade relationships between economies, where $\mathcal{E} = \{e_{ij}\}$ is the set of links $e_{ij}$. Figure 2 illustrates the size evolution of the four iCTNs from 1993–2018.

The number $N_\mathcal{V}$ of nodes involved in Figure 2A is the number of nodes, where $\mathcal{V}$ is the set of nodes. Compared with the iCTNs in 1993, the number of nodes of the maize and soybean networks increased markedly. It is consistent with previous literature [21]. However, the number of nodes of the rice and wheat networks kept stable with some slight fluctuations. Figure 2B shows the evolution of the number of links $N_\mathcal{E}$, which show excellent linear growth with respect to time *t*:

$$N_\mathcal{E}^{crop} = a^{crop} + b^{crop} t, \qquad (1)$$

A linear regression gives that $b^M = 50.09$ for maize, $b^R = 65.85$ for rice, $b^S = 27.90$ for soybean, and $b^W = 34.64$ for wheat. This highlights the fact that the number of links approximately increased linearly year by year. In general, the size of four iCTNs has expanded from 1993 to 2018. The network of rice had the largest size, indicating that more rice trade relations have been established between economies. Similarly, for soybean trade, less trade links have been established between economies. The increase in trade and network complexity differ for different iCTNs.

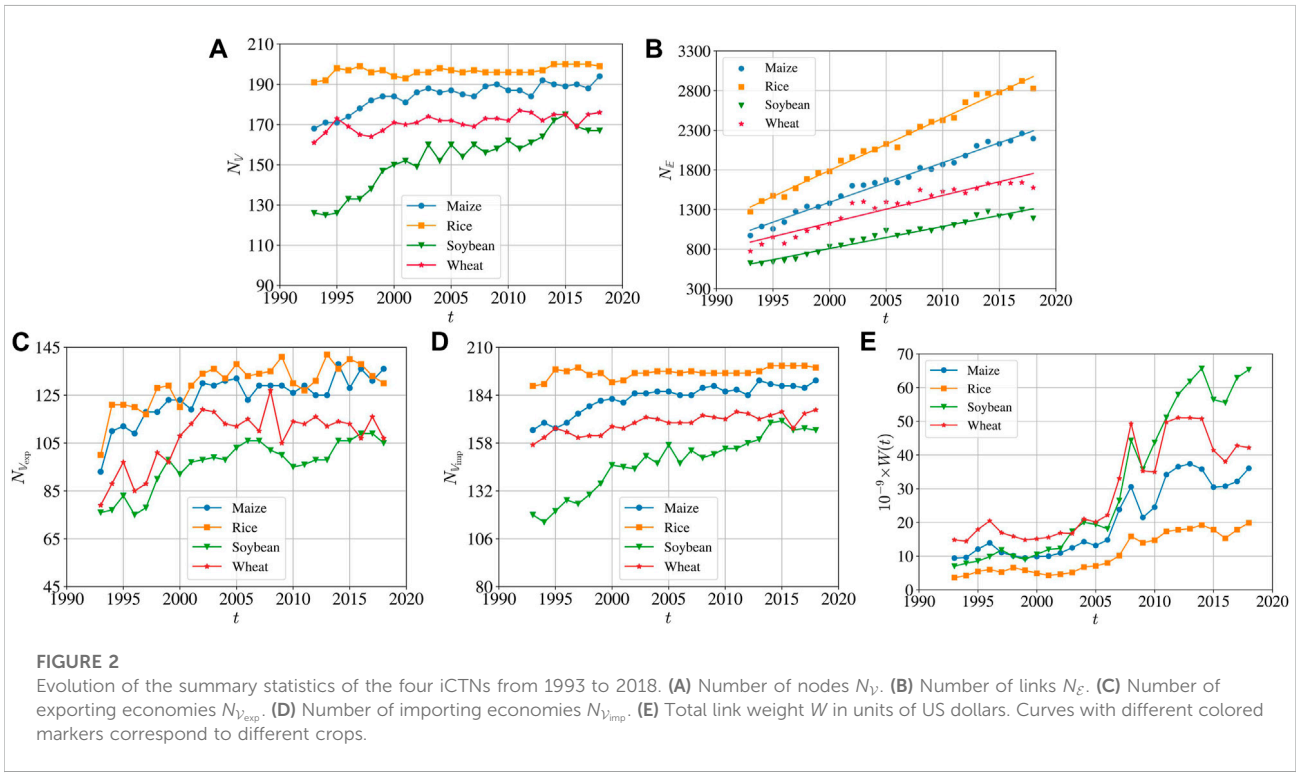

FIGURE 2
Evolution of the summary statistics of the four iCTNs from 1993 to 2018. **(A)** Number of nodes $N_\mathcal{V}$. **(B)** Number of links $N_\mathcal{E}$. **(C)** Number of exporting economies $N_{\mathcal{V}_\text{exp}}$. **(D)** Number of importing economies $N_{\mathcal{V}_\text{imp}}$. **(E)** Total link weight $W$ in units of US dollars. Curves with different colored markers correspond to different crops.

Figures 2C,D show the numbers of exporting and importing economies ($N_{\mathcal{V}_\text{exp}}$ and $N_{\mathcal{V}_\text{imp}}$) of the four iCTNs from 1993 to 2018. Compared with $N_\mathcal{V}$ in Figure 2A, we recognize that:

$$N_{\mathcal{V}_\text{exp}} < N_{\mathcal{V}_\text{imp}} < N_\mathcal{V} < N_{\mathcal{V}_\text{exp}} + N_{\mathcal{V}_\text{imp}}. \qquad (2)$$

There are much more importing economies than exporting economies, and many economies both export and import the same crops, which is also observed for the international pesticide trade networks [25].

Link weight presents the trade value between two economies. We calculated the sum of link weights to show the total trade value of a given crop. Figure 2E describes the evolution of the international trade values $W(t)$ of the four crops from 1993 to 2018. We find that the trade values of the four crops have an overall increasing trend, but decreased locally. Remarkably, $W(t)$ increased sharply in 2007/2008 due to the 2008 food crisis, which results in a significant rise in food prices and food insecurity [26]. Contributing factors are various, and macro-level underlying causes include higher oil prices, which affect the costs for food production and processing. Indeed, the oil market crashed in the middle of 2008 [27], followed by the prices of agricultural goods. In particular, a general rise in agricultural prices could create a global food price bubble [28]. Agricultural commodities exhibited unexpected price spikes again in 2011, prompting an increase in crop trade values [29]. Hence, we observe that $W(t)$ experienced a marked increase in 2011. In addition, rice had the lowest trade values in each year, and $W(t)$ of soybean overtook wheat to the highest in 2009.

## 3.2 Degree and strength

The node degrees show how many trade partners each economy has. In a directed network, we consider both in-degree and out-degree of a node to measure import and export respectively. The in-degree of node is defined as follows

$$k_i^\text{in} = \sum_{j \in \mathcal{V} - \{i\}} I_\mathcal{E}(e_{ji}) = \sum_{j=1}^{N_\mathcal{V}} I_\mathcal{E}(e_{ji}), \qquad (3)$$

where $I_\mathcal{E}(e_{ji})$ is the indicator function:

$$I_\mathcal{E}(e_{ji}) = \begin{cases} 1, & \text{if } e_{ji} \in \mathcal{E} \\ 0, & \text{if } e_{ji} \notin \mathcal{E} \end{cases} \qquad (4)$$

The out-degree of node is defined as follows

$$k_i^\text{out} = \sum_{j \in \mathcal{V} - \{i\}} I_\mathcal{E}(e_{ij}) = \sum_{j=1}^{N_\mathcal{V}} I_\mathcal{E}(e_{ij}). \qquad (5)$$

Since the networks are weighted, we quantity node strengths, including in-strength $s_i^\text{in}$ and out-strength $s_i^\text{out}$, which are defined as follows

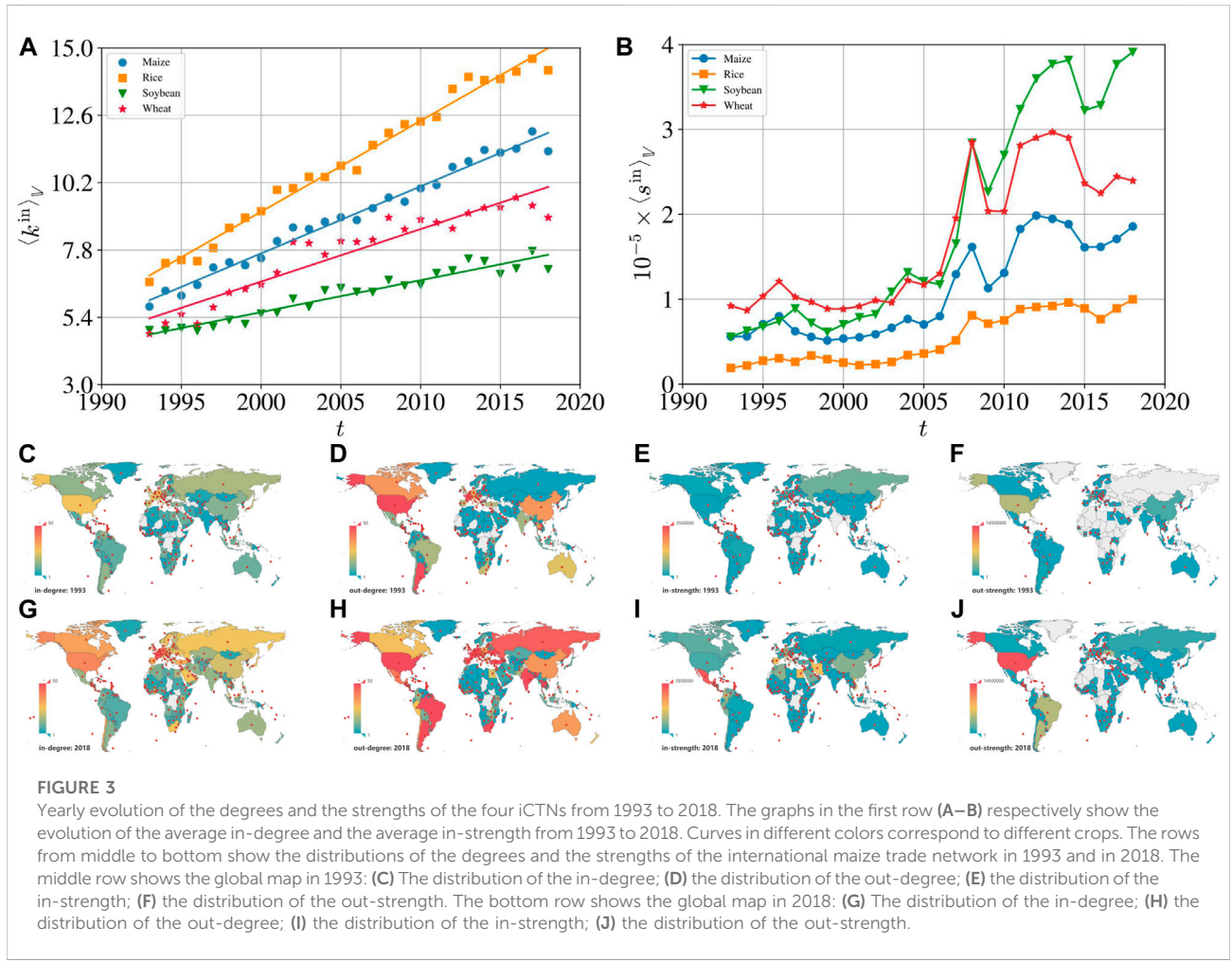

FIGURE 3
Yearly evolution of the degrees and the strengths of the four iCTNs from 1993 to 2018. The graphs in the first row **(A–B)** respectively show the evolution of the average in-degree and the average in-strength from 1993 to 2018. Curves in different colors correspond to different crops. The rows from middle to bottom show the distributions of the degrees and the strengths of the international maize trade network in 1993 and in 2018. The middle row shows the global map in 1993: **(C)** The distribution of the in-degree; **(D)** the distribution of the out-degree; **(E)** the distribution of the in-strength; **(F)** the distribution of the out-strength. The bottom row shows the global map in 2018: **(G)** The distribution of the in-degree; **(H)** the distribution of the out-degree; **(I)** the distribution of the in-strength; **(J)** the distribution of the out-strength.

$$s_i^{\text{in}} = \sum_{j \in \mathcal{V}-\{i\}} w_{ji} = \sum_{j=1}^{N_\mathcal{V}} w_{ji}, \quad (6)$$

$$s_i^{\text{out}} = \sum_{j \in \mathcal{V}-\{i\}} w_{ij} = \sum_{i=1}^{N_\mathcal{V}} w_{ij}, \quad (7)$$

where $w_{jj} = 0$ by definition.

The degrees and the strengths measure the importance of a node, and we used the average degrees and the average strengths to evaluate the overall structure of the networks. It is easy to get that the average in-degree of nodes $\langle k^{\text{in}} \rangle_\mathcal{V}$ is equal to the average out-degree of nodes $\langle k^{\text{out}} \rangle_\mathcal{V}$ [25]. The average in-strength $\langle s^{\text{in}} \rangle_\mathcal{V}$ and the average out-strength $\langle s^{\text{out}} \rangle_\mathcal{V}$ respectively measure the average values of imports and exports, which are also equal to each other.

Figures 3A,B show the yearly evolution of the average in-degree and in-strength from 1993 to 2018. The average node in-degree represents the average number of exporting partners owned to an economy [30], which is equal to the average node our-degree [25]. Figure 3A shows that the evolution of the average in-degree has an excellent linear growth with respect to time $t$:

$$k_\mathcal{V}^{\text{in, crop}} = a^{crop} + b^{crop} t. \quad (8)$$

Simple linear regressions give that $b^M = 0.24$ for maize, $b^R = 0.32$ for rice, $b^S = 0.11$ for soybean, and $b^W = 0.19$ for wheat. For all the iCTNs, the average in-degree increases with time, which indicates increasing active trade relationships among economies. The increasing trend of $k_\mathcal{V}^{\text{in}}$ is a result of the growth rate of $N_\mathcal{E}$ being greater than that of $N_\mathcal{V}$. We plotted the distributions of the in- and out-degree of the iCTNs in 1993 and 2018 to describe the explicit change. Since the results are similar in different iCTNs, here we only show the global maps of the international maize trade network. As shown in Figures 3C,D,G,H, both for in-degree and out-degree, the color of maps in 2018 was deeper than that in 1993, indicating an increase in the number of crop trade connections. However, $k_\mathcal{V}^{\text{in}}$ decreased markedly in 2018. It may be due to the fact that the growth of networks is not caused by simply adding new links to existing nodes which would disappear while new nodes are created [31].

As shown in Figure 3B, similar to the dynamics of link weights, the average in-strength increased across the sample and showed a potential upward trend. Before 2007, $s_\mathcal{V}^{in}$ kept a slight increase for each crop and occurred small fluctuations in some years. The average in-strength showed significant fluctuations in 1996–1997 and 2004–2005. The main factor that caused the average trade values to change dramatically is the food price. Prices for most crops started to climb slowly in 1990 and peaked in 1996 (maize, rice and wheat) and 1997 (soybean) before declining sharply. But financial crisis of 1997–99 quickly ended the crop price surge [32]. In 2004, due to bad harvests and high oil prices, the food prices increased, which caused an increase in global food trade values. The food prices slowed down as the global commodity prices were under control in 2005. The "world food crisis" of 2007–2008 inflated food prices significantly. This crisis originated from the long-term cycle of fossil-fuel dependence on industrial capitalism, coupled with the inflationary effect of current biofuel offset and financial speculation [33]. Under the influence of the food price crisis [34], the average trade values showed a significant upward trend from 2007 to 2008. After 2008, $s_\mathcal{V}^{in}$ reverted to increase with fluctuations. Overall, both for in-strength and out-strength, the color of maps in 2018 was deeper than that in 1993 as shown in Figures 3E,F,I,J, suggesting an increase in crop trade volumes.

## 3.3 Competitiveness

The density of a directed network refers to the ratio of the number of links that actually exist in the network to the number of all possible links:

$$\rho = \frac{N_\mathcal{E}}{N_\mathcal{V}(N_\mathcal{V} - 1)}, \quad (9)$$

To capture the potential relations between an economy's trading partners, we used the clustering coefficient to measure the connectivity of the economy's trading partners [35]. For a weighted network, the clustering coefficient of a node $i$ is the ratio of all directed triangles to all possible triangles [36],

$$c_i = \frac{2T_i}{k_i(k_i - 1) - 2k_i^R}, \quad (10)$$

where $T_i$ is the number of directed triangles containing node $i$, $k_i$ is the total degree of node $i$, and

$$k_i^R = \sharp(\{j: e_{ij} \in \mathcal{E} \,\&\, e_{ji} \in \mathcal{E}\}) = \sum_{j \neq i}(w_{ij}w_{ji})^0, \text{ for } w_{ij}w_{ji} \neq 0 \quad (11)$$

is the reciprocal degree of node $i$. We use the average clustering coefficient to measure the overall concentration of the network:

$$\langle c \rangle_\mathcal{V} = \frac{1}{N_\mathcal{V}} \sum_{i=1}^{N_\mathcal{V}} c_i. \quad (12)$$

Figure 4 illustrates the density and average clustering coefficient of the four crop trade networks, introduced to describe the competitiveness of the entire network. The value of density represents the tightness of a network [37]. In a dense network, the number of connections approaches to the maximum number of potential ties. According to Figure 4A, the density of each crop network was small. Over the last 26 years, the density rose with some fluctuations, and it indicates that the global food trade is becoming more and more frequent and close. The increasing densities of the iCTNs are consistent with some other international trade networks [38], but are much smaller [25, 39] than the total world trade networks. The density curves for rice, maize and wheat showed a linear upward trend, where the density of the rice network had the largest slope of increase and has become the largest since 1998. Although the number of links for the rice trade network increased significantly, its density did not change dramatically after 2012. From 2009 to 2012, the network density of the soybean trade network changed slightly without a dramatic trend, and fluctuated significantly after 2012.

The average clustering coefficient measures the overall concentration of connections in the network. Figure 4B shows that the economies were inclined to cluster together in the four iCTNs. The clustering coefficients have an upward trend, especially for the rice and soybean trade networks, which is consistent with conclusions from previous literature [40, 41]. Likewise, the values of clustering coefficients for rice were the largest after 2001 and displayed relatively least fluctuations, since the export and import of rice concentrated in some economies [42]. Compared with the evolution of the network density, a particularly dense network was inclined to have high clustering because its modes are more likely to share partners [38].

## 3.4 Persistence

There are two types of complex networks: multi-layer networks, in which nodes are connected in different ways; and temporal networks, in which nodes and links may appear or disappear and their attributes to the networks might change over time [43]. Node similarity has been widely studied for simple networks [44, 45] and multi-layer networks [43, 46]. This paper adopts a simple indicator to measure the similarity coefficient between two successive networks [25], since the links in iCTNs might look similar or change significantly. Considering two successive networks $\mathcal{G}(t-1)$ and $\mathcal{G}(t)$, let $\mathcal{E}_{(t-1)\cup t} = \mathcal{E}(t-1) \cup \mathcal{E}(t)$ be the union set of directed links and $\mathcal{E}_{(t-1)\cap t} = \mathcal{E}(t-1) \cap \mathcal{E}(t)$ be the intersection of directed links. Based on previous studies [25, 47, 48], we define the temporal similarity between two successive networks $\mathcal{G}(t-1)$ and $\mathcal{G}(t)$ as the ratio of the number of overlapping directed links in the two networks over the number of all directed links in the two networks:

$$S(t) = \frac{\#(\mathcal{E}_{(t-1)\cap t})}{\#\mathcal{E}_{(t-1)\cup t}}. \qquad (13)$$

where $\#(\mathbf{X})$ denotes the cardinal number of set $\mathbf{X}$. The value of the similarity coefficient $S(t)$ ranges between 0 and 1: $S(t) = 0$ indicates that the two networks are completely different in means of links, while $S(t) = 1$ means that the two networks are completely the same.

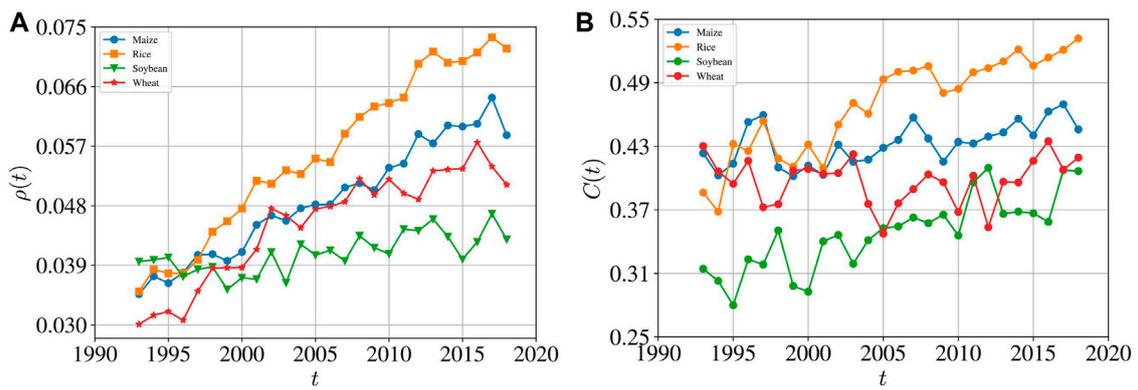

**FIGURE 4**
Yearly evolution of the network density **(A)** and average clustering coefficient **(B)** of the four international crop trade networks from 1993 to 2018.

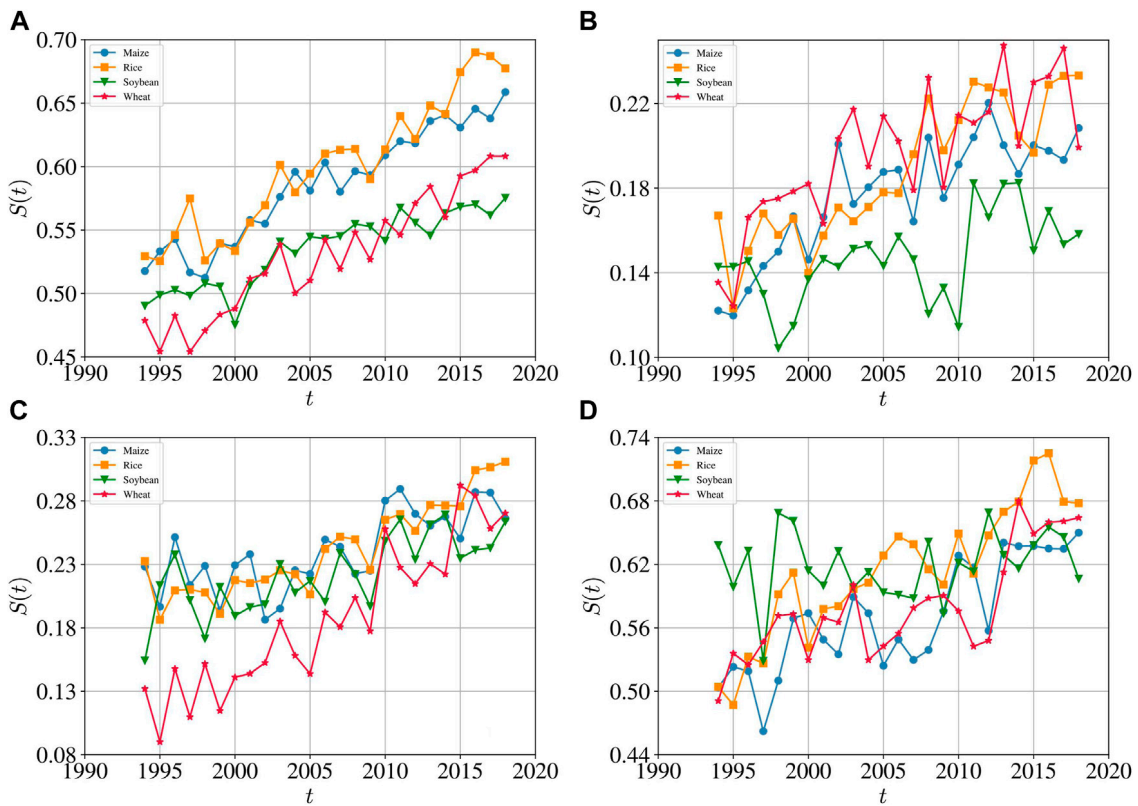

**FIGURE 5**
Evolution of the temporal similarity coefficient $S(t)$ between two successive networks of the four crops from 1993 to 2018. **(A)** All links at each time. **(B)** Light links with the weights at each time less than the 20% percentile. **(C)** Medium links with the weights at each time between the 40 and 60% percentiles. **(D)** Heavy links with the weights at each time greater than the 80% percentile. The temporal similarity increased over time with slight fluctuations.

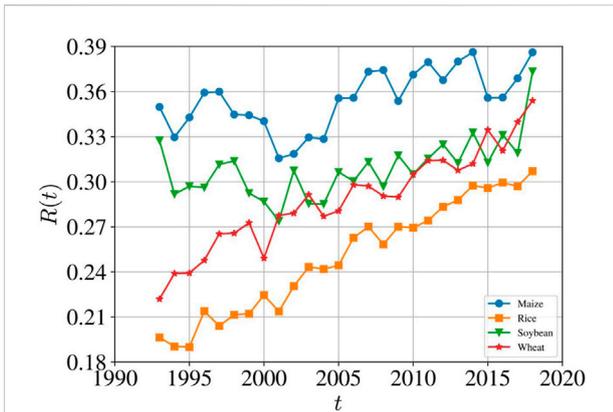

FIGURE 6
Evolution of overall reciprocity of the four iCTNs from 1993 to 2018. The overall reciprocity coefficients were between 0.1 and 0.4.

The analysis of the node similarity is a significant basis for understanding the evolution of features of the international crop trade system. The small value of the similarity coefficient $S$ shows a high discrepancy in the structure of two successive networks [25, 49]. From Figure 5A, the temporal similarity increased over time with slight fluctuations, which indicates that the structure of successive iCTNs gets more similar. And the rice trade network had the largest temporal similarity recently. By comparing the similarity coefficient $S(t)$ of sub-networks containing links with different values of weight (light links in Figure 5B where the weights are less than the 20% percentile, medium links in Figure 5C where the weights are between the 40 and 60% percentiles, and heavy links in Figure 5D where the weights are greater than the 80% percentile), it can be found that the $S(t)$ curves have similar patterns qualitatively and the heavier links with greater trade flows have more stable.

## 3.5 Reciprocity

The reciprocity is critical to dynamical processes and network growth [50]. The reciprocity of a directed network is defined as the ratio of the number of bilateral links (i.e., links pointing in both directions) to the total number of links in the network [51, 52]:

$$R = \frac{\#(\{(i,j): e_{ij} \in \mathcal{E} \ \& \ e_{ji} \in \mathcal{E}\})}{\#(\{(i,j): e_{ij} \in \mathcal{E}\})} = \frac{1}{N_{\mathcal{E}}} \sum_{i \in V} k_i^R, \qquad (14)$$

where

$$\#(\{(i,j): e_{ij} \in \mathcal{E} \ \& \ e_{ji} \in \mathcal{E}\}) = \sum_{i \in V} k_i^R \qquad (15)$$

and

$$\#(\{(i,j): e_{ij} \in \mathcal{E}\}) = N_{\mathcal{E}}. \qquad (16)$$

Reciprocity $R$ is an indicator of the degree of bilateral trade relationships between economies in a network and plays an important role in the transmission mechanism of international trade information.

Figure 6 shows the evolution of overall reciprocity of the four iCTNs from 1993 to 2018. The overall reciprocity coefficients were between 0.1 and 0.4. It is found that the overall reciprocity was relatively stable with slight fluctuations for maize and soybean. In terms of wheat and rice, the reciprocity coefficients were always smaller than those of the maize and soybean trade networks, but showed an increasing trend. Especially for rice, the reciprocal coefficient $R(t)$ showed a nice linear relationship with time $t$. We note that the reciprocity coefficients of the iCTNs are much smaller than those of the international trade networks (larger than 0.5) [39, 51], which contain remarkably more commodities and thus more reciprocal links.

## 3.6 Assortativity

Assortativity quantifies the mixing pattern of complex networks, which measures whether the node is preferentially connected to a node with a similar scale [53]. In a directed network, we consider the correlation of four degree directions. The degree assortative coefficient $r_{\text{in,in}}(t)$ between the in-degree of exporting economies and the in-degree of importing economies:

$$r_{\text{in,in}}(t) = \frac{1}{N_{\mathcal{E}}} \sum_{e_{ij} \in \mathcal{E}} \frac{\left[\left(k_i^{\text{in}} - \langle k_i^{\text{in}} \rangle_{\mathcal{E}}\right)\left(k_j^{\text{in}} - \langle k_j^{\text{in}} \rangle_{\mathcal{E}}\right)\right]}{\sigma_{i,\mathcal{E}}^{\text{in}} \sigma_{j,\mathcal{E}}^{\text{in}}}, \qquad (17)$$

where $\langle k_i^{\text{in}} \rangle_{\mathcal{E}}$ and $\langle k_j^{\text{in}} \rangle_{\mathcal{E}}$ are respectively the mean in-degrees of exporting economies and importing economies, and the variance of in-degrees of exporting economies is

$$\left(\sigma_{i,\mathcal{E}}^{\text{in}}\right)^2 = \frac{1}{N_{\mathcal{E}}} \sum_{e_{ij} \in \mathcal{E}} \left(k_i^{\text{in}} - \langle k_i^{\text{in}} \rangle_{\mathcal{E}}\right)^2, \qquad (18)$$

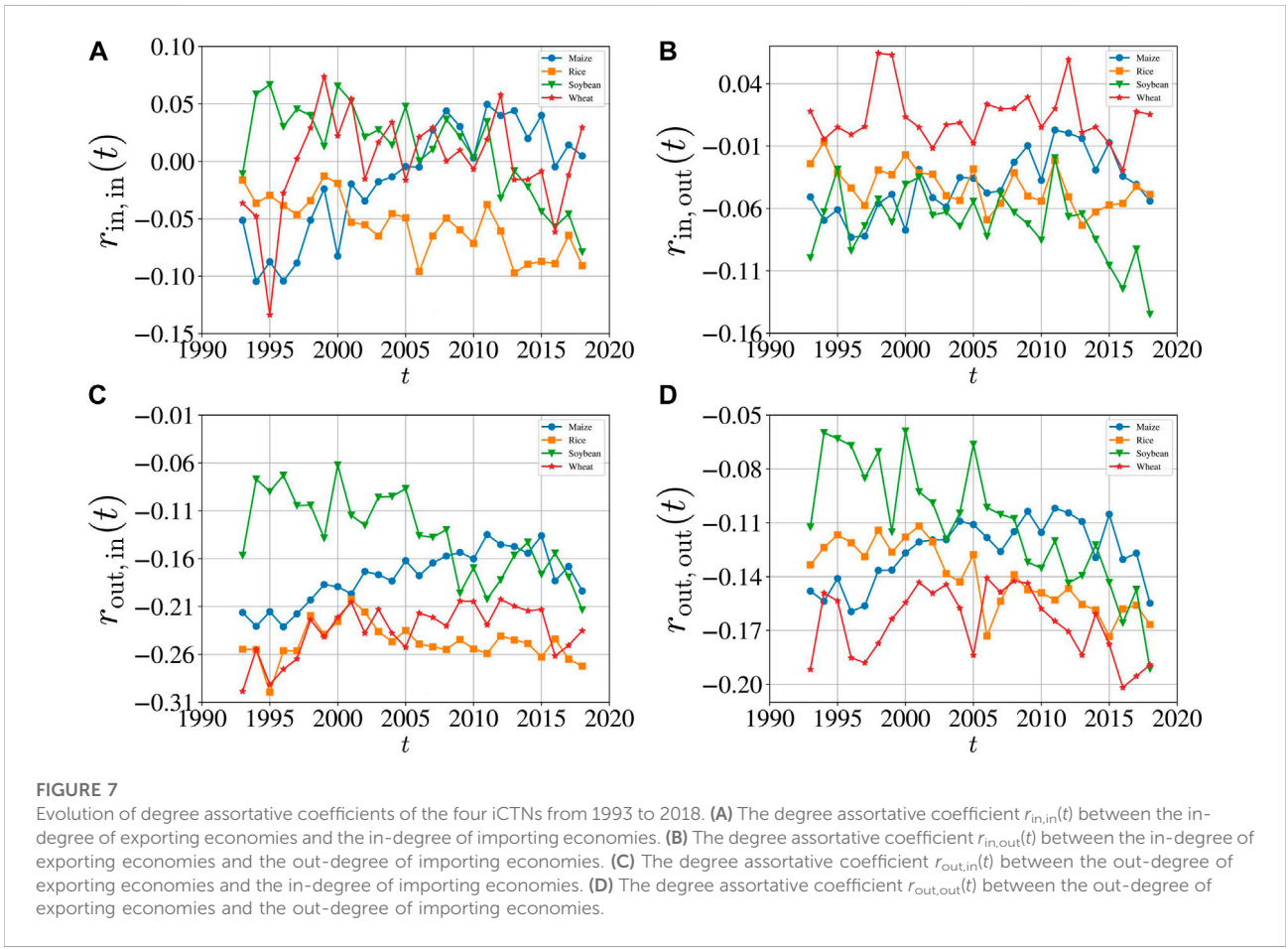

FIGURE 7
Evolution of degree assortative coefficients of the four iCTNs from 1993 to 2018. **(A)** The degree assortative coefficient $r_{in,in}(t)$ between the in-degree of exporting economies and the in-degree of importing economies. **(B)** The degree assortative coefficient $r_{in,out}(t)$ between the in-degree of exporting economies and the out-degree of importing economies. **(C)** The degree assortative coefficient $r_{out,in}(t)$ between the out-degree of exporting economies and the in-degree of importing economies. **(D)** The degree assortative coefficient $r_{out,out}(t)$ between the out-degree of exporting economies and the out-degree of importing economies.

and the variance $(\sigma_{j,\mathcal{E}}^{in})^2$ of in-degrees of importing economies is defined in the same way.

Similarly, the degree assortative coefficient $r_{in,out}(t)$ between the in-degree of exporting economies and the out-degree of importing economies is

$$r_{in,out}(t) = \frac{1}{N_\mathcal{E}} \sum_{e_{ij} \in \mathcal{E}} \frac{\left[\left(k_i^{in} - \langle k_i^{in} \rangle_\mathcal{E}\right)\left(k_j^{out} - \langle k_j^{out} \rangle_\mathcal{E}\right)\right]}{\sigma_{i,\mathcal{E}}^{in} \sigma_{j,\mathcal{E}}^{out}}, \quad (19)$$

the degree assortative coefficient $r_{out,in}(t)$ between the out-degree of exporting economies and the in-degree of importing economies is

$$r_{out,in}(t) = \frac{1}{N_\mathcal{E}} \sum_{e_{ij} \in \mathcal{E}} \frac{\left[\left(k_i^{out} - \langle k_i^{out} \rangle_\mathcal{E}\right)\left(k_j^{in} - \langle k_j^{in} \rangle_\mathcal{E}\right)\right]}{\sigma_{i,\mathcal{E}}^{out} \sigma_{j,\mathcal{E}}^{in}}, \quad (20)$$

and the degree assortative coefficient $r_{out,out}(t)$ between the out-degree of exporting economies and the out-degree of importing economies is

$$r_{out,out}(t) = \frac{1}{N_\mathcal{E}} \sum_{e_{ij} \in \mathcal{E}} \frac{\left[\left(k_i^{out} - \langle k_i^{out} \rangle_\mathcal{E}\right)\left(k_j^{out} - \langle k_j^{out} \rangle_\mathcal{E}\right)\right]}{\sigma_{i,\mathcal{E}}^{out} \sigma_{j,\mathcal{E}}^{out}}. \quad (21)$$

The four degree assortative coefficients of different directions can be used to describe the relevance of two nodes connected by a directed link through their in- and out-degrees to accurately explore the mixing patterns of the international crop trade networks.

Figure 7 shows the evolution of the degree assortative coefficients of the four iCTNs from 1993 to 2018. The $r$ values fluctuated sharply before 2000, followed by relatively mild fluctuations. Previous research that did not consider the direction of the degree has shown that world cereal trade networks [54] or other international trade networks [39, 55] are disassortative. In this paper we find that the degree assortative coefficients of different directions for different crop networks have different assortative patterns. As shown in Figure 7A, the degree assortative coefficients $r_{in,in}(t)$ ranged from −0.2 to 0.1. For maize, the coefficients were almost negative before 2006, and showed an upward trend until 2015. For rice, the coefficients were always negative. For soybean, the coefficients showed significant fluctuations before 1994, ranged from 0 to 0.1 with fluctuations during 1995–2011, and finally dropped to less than zero. From Figure 7B, the coefficients $r_{in,out}(t)$ for maize, rice and soybean were almost negative, while the coefficients for wheat were mainly positive. According to Figures 7C,D, for all the iCTNs, the degree assortative coefficients $r_{\text{out,in}}(t)$ and $r_{\text{out,out}}(t)$ were generally negative. In summary, except that almost all the assortative coefficients for the international rice trade network were negative, the iCTNs exhibit complex mixing patterns.

## 4 Conclusion

Achieving global food security is one of the major challenges of the coming decades [56], and network analysis has been a popular approach to understanding the international food trade system. In this paper, we focused on four important crops (maize, rice, soybean and wheat), and provided a time series analysis of the four international crop trade networks from 1993 to 2018. Rather than investigating one multiplex trade network *via* combining several goods, we analyzed the international trade networks of individual crops and carried out comparisons. We revealed the evolution of topological properties, including degrees, strengths, link weights, density, clustering coefficient, reciprocity, and assortativity.

We found that the sizes of all the four iCTNs expanded from 1993 to 2018 with more involved international trading participants and larger trade values. The number of links also significantly increased, indicating that many new trade relationships were formed in the global food trade system over the past decades. The link weights decreased sometimes, but showed an increasing trend in general for the four crops. As the networks are directed, we calculated the in-degree, out-degree, in-strength and out-strength to explicitly understand the trade flow in the global food system. The average in- and out-degree increased, representing a larger number of active trade relationships among economies. The increasing trade partnerships, network density, clustering coefficients and similarity coefficients consistently witness the globalization of the international crop trade.

We found that the density of each crop network was low. Over the last 26 years, the density rose with local fluctuations. Our findings are consistent with some other international trade networks [38], but are much smaller [25, 39] than the total world trade networks. The clustering coefficients also showed an upward trend, especially for the rice and soybean trade networks. The structure of the iCTNs become not only tighter but also more similar. In addition, the networks with greater trade flows have more stable relationships. In each iCTN, the reciprocity coefficients were between 0.1 and 0.4, and much smaller than those of the international trade networks. We also obtained some interesting results. For example, although most iCTNs were disassortatively mixed, there were iCTNs exhibiting assortative mixing patterns in certain years, which unveils more complicated mixing behavior than an overall assortative coefficient for the world cereal trade networks [54] or other international trade networks [39, 55].

We compared the structure of four iCTNs to enhance our understanding of the global food system. Although the overall evolution of different iCTNs is similar, some crops have unique trade patterns. For example, the average in-degree of the international wheat trade network decreased in 2011, contrary to other crops. It might be affected by the Russian wheat export ban in 2010–2011, which caused a decrease in the trade flow [57]. The density of the international rice trade network has the largest increase and has become the largest since 1998. The evolution of the clustering coefficients shows that the international rice trade network became more clustered, since the rice exporting and importing concentrated in some economies [42].

Our findings about the topology of the iCTNs play an important role in understanding vulnerabilities in the global food system [11]. These results also highlight the need to consider unique features of different crop networks while making food policies [11]. Since each iCTN has its own structural properties, they are expected to have different reactions to external disturbances and shocks. The global food system is sensitive and easily affected by climate change, water scarcity, and land reclamation [58]. For example, we could assume that an extreme climate decreases the production of crops in some areas which are main global crop suppliers. These economies would cut down crop exports and even implement export bans if their domestic food reserves are insufficient. However, we found that the density of the international rice trade network showed an upward trend during the recent food crisis (e.g., in 2007–2009). As the international rice trade network is increasingly connected, the rice trade tends to concentrate on some regions. A few large producers account for the bulk of net exports and absorb more shocks because of their centrality in the network [59]. These economies are not sensitive to global changes since they have proportionately higher reserves. Therefore the international rice trade network is relatively stable and its structure would not shift dramatically.

In addition to environmental factors, global price shocks also exert a significant influence on the global food system [60], especially for rice, the main staple crop. Many economies rely on rice imports to feed domestic consumption and the rice price hike would put more pressure on importing economies [61], limiting the poor to buying rice [2]. Demand for substitute staple foods increases to soften the impact of rice price shocks [61]. The iCTNs are characterized by substantial heterogeneity across different crops, but crops are traded as complements which indicates that different iCTNs might have a correlation [11]. This paper discussed the global food system as a collection of independent food-staple trade players, and ignored the substitution across crops. However, our findings are still relevant from a policy perspective. As noted above, the similarities and differences between different iCTNs provide more details of the global food trade linkages and address the need to adjust trade policies for different crop importers or exporters. Future research should consider the nonlinear interactions between different iCTNs from the framework of multi-layer networks.

## Data availability statement

Publicly available datasets were analyzed in this study, which can be found here: https://www.fao.org.

## Author contributions

Funding acquisition, W-XZ; Investigation, Y-TZ; Methodology, Y-TZ and W-XZ; Supervision, W-XZ; Writing—original draft, Y-TZ and W-XZ; Writing—review and editing, Y-TZ, and W-XZ.

## Funding

The National Natural Science Foundation of China (72171083), the Shanghai Outstanding Academic Leaders Plan, and the Fundamental Research Funds for the Central Universities.

## Conflict of interest

The authors declare that the research was conducted in the absence of any commercial or financial relationships that could be construed as a potential conflict of interest.

## Publisher's note

All claims expressed in this article are solely those of the authors and do not necessarily represent those of their affiliated organizations, or those of the publisher, the editors and the reviewers. Any product that may be evaluated in this article, or claim that may be made by its manufacturer, is not guaranteed or endorsed by the publisher.